\documentclass[%
 aip,
 jcp,
 amsmath,amssymb,
 reprint,% use this for corrections
]{revtex4-1}
\usepackage{xcolor}
\usepackage{physics}
\usepackage{subcaption}
\usepackage{siunitx}
\usepackage{hyperref}
\usepackage{dirtytalk}
\usepackage{amssymb}
\usepackage{mathrsfs}

\usepackage{amsmath}
\usepackage{expl3}
\usepackage{calc}
\usepackage[version=4]{mhchem}

\usepackage{upgreek}

\usepackage{graphicx}% Include figure files
\usepackage{dcolumn}% Align table columns on decimal point
\usepackage{bm}% bold math
\usepackage[utf8]{inputenc}
\usepackage[T1]{fontenc}
\usepackage{mathptmx}
\usepackage{etoolbox}

\usepackage{booktabs}

\usepackage{enumitem}
\newlist{todolist}{itemize}{2}
\setlist[todolist]{label=$\square$}
\usepackage{pifont}
\makeatletter
\def\@email#1#2{%
 \endgroup
 \patchcmd{\titleblock@produce}
  {\frontmatter@RRAPformat}
  {\frontmatter@RRAPformat{\produce@RRAP{*#1\href{mailto:#2}{#2}}}\frontmatter@RRAPformat}
  {}{}
}%
\makeatother

\begin{document}

\preprint{AIP/123-QED}

\title[Intramolecular nuclear dynamics in ICEC]{Intramolecular nuclear dynamics in intermolecular Coulombic electron capture}

% Force line breaks with \\



%\author{}
%\thanks{}
%\affiliation{}

\author{Elena M. Jahr}
\email{elena.jahr@uni-tuebingen.de}
    \affiliation{Institute of Physical and Theoretical Chemistry, University of Tübingen, Auf der Morgenstelle 18, 72076 Tübingen, Germany}
    \affiliation{Center for Light-Matter Interaction, Sensors and Analytics (LISA+), University of Tübingen, Auf der Morgenstelle 15, 72076 Tübingen, Germany}

\author{Elke Fasshauer}
    \email{elke.fasshauer@pm.me}
    \affiliation{Institute of Physical and Theoretical Chemistry, University of Tübingen, Auf der Morgenstelle 18, 72076 Tübingen, Germany}
    \affiliation{Center for Light-Matter Interaction, Sensors and Analytics (LISA+), University of Tübingen, Auf der Morgenstelle 15, 72076 Tübingen, Germany}


\date{\today}% It is always \today, today,

\begin{abstract} 
We present an analytical model for intermolecular Coulombic electron capture (ICEC) which includes the internal nuclear dynamics of the molecules involved.
In ICEC, an electron attaches to an atom or molecule by transferring excess energy to a neighbor, ionizing said neighbor.
While previous theoretical investigations assumed fixed nuclei, recent studies indicate that relative motion between the two ICEC partners significantly influences the process.
Here, we incorporate the internal nuclear motion of the molecules involved into an analytical equation of the ICEC cross section.
We employ two approaches:
1. utilizing theoretical vibrationally resolved photoionization cross sections
and 2. applying the Franck-Condon principle.
Our theory yields electron spectra, ICEC cross sections for individual vibronic transitions, and temperature dependent cross sections.
Nuclear dynamics lead to a distribution of the electronic cross section over several vibrational states
and, in our model system \ce{H+} LiH, triggers dissociation of LiH during ICEC.

\end{abstract}


\maketitle %\maketitle must follow title, authors, abstract and \pacs

\section{Introduction}

Interparticle Coulombic Electron Capture (ICEC) is a non-radiative electron capture (or attachment) mechanism predicted to occur in atoms, molecules, or quantum dots embedded in an environment~\cite{gokhberg2009, gokhberg2010, bande_2023}.
In ICEC, a free electron attaches to an electron acceptor (A), transferring its excess energy to a neighboring electron donor (D), resulting in the ionization of D:
\begin{equation}
\label{eq:icec}
e^-_k + \mathrm{A} + \mathrm{D} \to \mathrm{A}^- + \mathrm{D}^+ + e^-_{k'}.
\end{equation}
ICEC can be described as an electron scattering process and its efficiency can thus be quantified with the scattering cross section\cite{gokhberg2009,gokhberg2010}.

In general, the capture of free electrons is a fundamental process in both basic and applied sciences. 
It plays a key role in astrophysics and plasma physics, where it influences stellar processes and ionized gases~\cite{tucker, astro}. 
Furthermore, electron attachment is critical for understanding electrochemical and photochemical reactions, impacting biological systems~\cite{dna} and solid-state technologies~\cite{semiconductors}. 
ICEC is therefore expected to be relevant in these
scenarios as well.

Various theoretical methods have been extended to study the different ICEC pathways in systems such as \ce{NeHe+}\cite{sisourat_2018}, \ce{H+} \ce{H2O}\cite{molle_2021}, and quantum dots\cite{pont_2013}, 
demonstrating that ICEC enhances electron attachment probability in weakly interacting systems compared to isolated photorecombination, or radiative attachment for a neutral atom.
A comprehensive review on ICEC, including schematic illustration of its different variants, can be found in Ref.~\onlinecite{bande_2023}.
Although ICEC has yet to be observed in experiment; efforts are underway to achieve the first measurement.

\begin{figure}[!ht]
    \centering
    \includegraphics[width=0.9\linewidth]{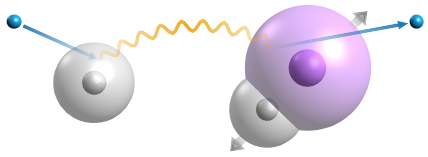}
    \caption{Schematic picture of interparticle Coulombic electron capture: A free electron attaches to an atom or molecule by transferring excess energy to a neighbor.
    If one unit is a molecule, their electron attachment or detachment process, respectively, is coupled to nuclear degrees of freedom.
    Figure taken from Ref.~\onlinecite{figshare_figures}.
    }
    \label{fig:icec}
\end{figure}

While ICEC has been extensively studied for atoms and for small molecules under the assumption of static nuclei\cite{bande_2023}, 
the effects of nuclear degrees of freedom were not taken into account until recently.
We extended the asymptotic model\cite{gokhberg2010} for the direct ICEC \cite{jahr2025} and for the related two-center dielectronic recombination \cite{jan2025} to include nuclear dynamics between the electron acceptor and donor.
Our findings, in conjunction with earlier studies \cite{fede2024}, reveal that interparticle nuclear motion significantly affects the ICEC cross sections.
The coupling to nuclear degrees of freedom allows ICEC below the vertical energy threshold\cite{fede2024, jahr2025} and can cause the dissociation of the weakly bound system\cite{jahr2025}.  
Inclusion of vibrational states further leads to the distribution of the purely electronic cross section over multiple final vibrational states, resulting in modulated outgoing electron energies and a broadened ICEC electron spectrum\cite{jahr2025}.
However, it is expected that not only the investigated \emph{inter}particle nuclear dynamics, but also the \emph{intra}particle nuclear dynamics, illustrated in Fig.~\ref{fig:icec}, will affect key properties of the ICEC process.

The relevance of intramolecular nuclear motion on similar energy transfer process has previously been studied.
For example, vibrational ICD -- vibrational de-excitation of a molecule with simultaneous ionization of a neighbor --
can decrease the lifetime of a vibrationally excited molecule from (milli)seconds to nanoseconds and picoseconds.\cite{cederbaum_ultrafast_2018}
It was further shown in 2020 that a molecule can dissociate following an energy transferred from a remote neighbor\cite{cederbaum_2020}.

In this work, we focus on these intramolecular nuclear dynamics of the electron acceptor and donor within the direct ICEC process.
We base our derivation on the asymptotic approximation which assumes weak interaction and negligible overlap between the wavefunctions of the two ICEC partners\cite{gokhberg2010}.
By extending this analytical model, we aim to bridge the gap between static-nuclei models and the complex reality of molecular systems.

Although we include internal nuclear dynamics for both ICEC partners in the theory, we will apply the model to an atom together with a diatomic molecule.
This reduces the effect of nuclear motion to one ICEC partner, while diatomic molecules allow for a clear analysis of the interplay between internal nuclear motion and electronic transitions.
We choose the proton (\ce{H+}) to act as the electron acceptor and lithium hydride (LiH) as the electron donor, forming a weakly interacting system whose processes are important in early cosmochemistry\cite{Li_2003, dawn_of_chemistry, HLiH_2023}.
ICEC would allow a transition from \ce{H+} LiH to H \ce{LiH+} otherwise only possible by interaction with radiation\cite{HLiH_2023}.
Given the significance of LiH, previous studies provide theoretical vibrationally resolved photoionization cross sections\cite{lundsgaard_LiH}, allowing us to explore different levels of theory.

This paper provides a foundational framework for understanding ICEC in molecular systems, paving the way for future studies on clusters and extended materials.
It is organized as follows: 
Section~\ref{sec:theory} describes the theoretical approaches employed in this work, while Section~\ref{sec:systems} introduces \ce{H+} LiH and Section~\ref{sec:comp-details} outlines the computational details. 
In Section~\ref{sec:results}, we present and discuss our results, highlighting the effects of intramolecular nuclear dynamics on ICEC. 
Finally, we summarize our findings and their broader implications in Section~\ref{sec:conclusion}.
We use atomic units ($\hbar = m_e = e = 4\pi\epsilon_0 = 1$) throughout this paper, unless explicitly mentioned otherwise.

\section{Theory}
\label{sec:theory}

From a scattering theory perspective, the ICEC process given in Eq.~\eqref{eq:icec} is a multichannel inelastic process in which an incoming electron interacts with a target~\cite{gokhberg2009, taylor2006}.
The efficiency of ICEC is thus characterized by the differential cross section for an incoming and outgoing electron of momentum $\mathbf{k}$ and $\mathbf{k}'$, respectively, 
\begin{equation}
    \label{eq:diff_xs}
    \dv{\sigma}{\Omega_{\mathbf{k}'}} 
    = \frac{1}{(2\pi)^2}\frac{k'}{k} |M|^2.
\end{equation}
Here, $\Omega_{\mathbf{k}'}$ is the solid angle of the outgoing electron and $M$ is the transition matrix element~\cite{gokhberg2009, taylor2006}.

We assume A and D to be sufficiently separated for the electronic interaction between them to be weak, justifying the use of perturbation theory.
From a purely electronic perspective, the transition matrix element can then be written as 
\begin{equation}
\label{eq:M_electronic}
    M =  \langle \phi_{\mathrm{A},\mathbf{k}} \, \phi_\mathrm{D} | \hat{V}_C |\phi_{\mathrm{A}^-} \, \phi_{\mathrm{D}^+,\mathbf{k}'} \rangle,
\end{equation}
where $\phi_\mathrm{D}$ and $\phi_{\mathrm{A}^-}$ are electronic wavefunctions and $\hat{V}_C$ is the Coulomb interaction between the charges of A and D.
The momentum-normalized scattering state $\phi_{\mathrm{A}, \mathbf{k}}$ ($\phi_{\mathrm{D}^+, \mathbf{k'}}$) describes A (D$^+$) together with the incoming (outgoing) electron.\cite{gokhberg2009, gokhberg2010}

This approach is referred to as "asymptotic" or "virtual photon" approximation in the literature\cite{bande_2023} and leads to the electronic cross section
\begin{equation}
    \label{eq:icec_electronic}
    \sigma(\varepsilon) 
    = 
    \frac{3 c^4 }{4 \pi} \,
    \frac{
    \sigma^\mathrm{PR}_\mathrm{A}(\varepsilon) \; 
    \sigma^\mathrm{PI}_\mathrm{D}(\omega)
    }{\omega^4 \, R_\mathrm{AD}^6},
\end{equation}
where $\sigma^\mathrm{PR}_\mathrm{A}$ and $\sigma^\mathrm{PI}_\mathrm{D}$ are the photorecombination (PR) and -ionization (PI) cross sections of A and D, respectively.
For neutral A, this would correspond to radiative attachment or photoattachment, while for ionic D it would be photodetachment.
$R_\mathrm{AD}^6$ is the distance between the center of masses of A and D.
The transferred energy $\omega$ is the sum of the kinetic energy of the incoming electron $\varepsilon$ and the electron affinity of A, or equivalently the ionization potential (IP) of $\mathrm{A}^-$, $\omega = \epsilon + \mathrm{IP}_{\mathrm{A}^-}$.

This asymptotic approximation forms a lower bound to the electronic cross section of ICEC for small distances $R_\mathrm{AD}^6$ and increases in accuracy with increasing separation where the electronic interaction is weak\cite{bande_2023}.
For atomic ICEC partners, the energy transfer model in Eq.~\eqref{eq:icec_electronic} has been extended to include the electron transfer pathway at small distances into the analytical description of ICEC\cite{senk2024}.

We recently included the effect of nuclear motion between A and D in the description of both the energy and electron transfer mechanisms.\cite{jahr2025,jan2025}
In this work, we focus on extending the asymptotic approach to include the internal vibrational degrees of freedom of A and D into the energy transfer mechanism of ICEC.

\subsection{Asymptotic ICEC including internal vibrational dynamics}
\label{sec:theory_asymptotic}

We begin by writing the initial ($i$) and final ($f$) wave functions as products of the individual states of A and D\cite{gokhberg2009},
\begin{equation}
\label{eq:initial+final_states}
\begin{aligned}
    \Psi_i &= \psi_{\mathrm{A}, \mathbf{k}} \psi_\mathrm{D} \\
    \Psi_f &= \psi_{\mathrm{A}^-} \psi_{\mathrm{D}^+, \mathbf{k}'}. \\
\end{aligned}
\end{equation}
where $\psi_\mathrm{D}$ and $\psi_{\mathrm{A}^-}$ are now vibronic states.
The momentum-normalized scattering state $\psi_{\mathrm{A}, \mathbf{k}}$ ($\psi_{\mathrm{D}^+, \mathbf{k'}}$) describes molecule A (D$^+$) in a vibronic state together with the incoming (outgoing) electron.
The transition matrix element in Eq.~\eqref{eq:M_electronic} is then given as
\begin{equation}
\label{eq:M}
    M =  \langle \psi_{\mathrm{A}^-} \, \psi_{\mathrm{D}^+,\mathbf{k}'} | \hat{V}_C | \psi_{\mathrm{A},\mathbf{k}} \, \psi_\mathrm{D} \rangle.
\end{equation}

We describe the vibronic states of A and D in Eq.~\eqref{eq:initial+final_states} and \eqref{eq:M} as a single term of the Born-Huang expansion, \cite{Ballhausen1972, Born1996}
\begin{equation}
\label{eq:BH_expansion}
    |\psi_{\alpha} (\bm{r},\bm{R}) \rangle = |\phi_\alpha(\bm{r}\,;\bm{R})\rangle \, |\nu_\alpha(\bm{R})\rangle,
\end{equation}
where $\phi_\alpha$ is the electronic wave function, $\nu_\alpha$ is the nuclear wave function, $\bm{r}$ are the electronic coordinates, and $\bm{R}$ are the nuclear coordinates within a molecule.
Here, we neglect the interaction between vibrational and rotational degrees of freedom and assume the final vibrational states to be represented by those of the ions.

\begin{figure}[!ht]
    \centering
    \includegraphics[width=0.9\linewidth]{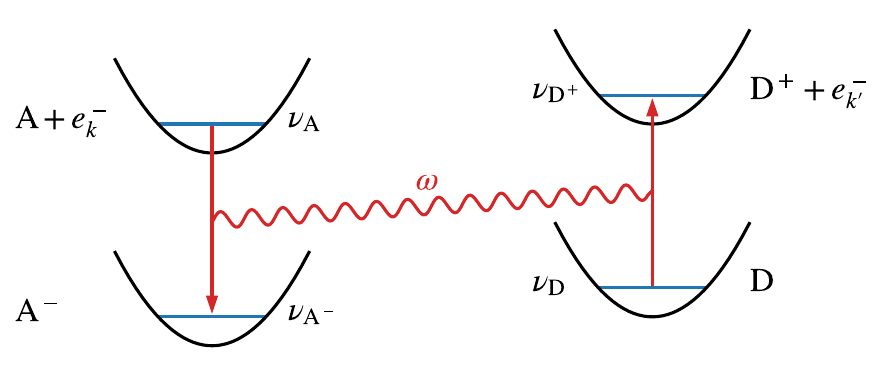}
    \caption{Energetic representation of ICEC with intramolecular nuclear dynamics:
    Molecule A binds a free electron of momentum $k$ and transitions from vibronic state $\phi_\mathrm{A}\nu_\mathrm{A}$ to $\phi_{\mathrm{A}^-}\nu_{\mathrm{A}^-}$. 
    Excess energy $\omega$ is transferred to molecule D, ionizing D from $\phi_\mathrm{D}\nu_\mathrm{D}$ to $\phi_{\mathrm{D}^+}\nu_{\mathrm{D}^+}$ with a continuum electron of momentum $k'$.
    Figure taken from Ref.~\onlinecite{figshare_figures}.
    }
    \label{fig:icec_sketch}
\end{figure}

The ICEC process we discuss is sketched in Fig.~\ref{fig:icec_sketch} and proceeds as follows:
Molecule A in a vibrational level $\nu_{\mathrm{A}}$ of the electronic ground state $\phi_\mathrm{A}$ captures an electron of kinetic energy $\varepsilon$, transitioning to the vibrational state $\nu_{\mathrm{A}^-}$ of the anionic ground state $\phi_{\mathrm{A}^-}$. 
The excess energy $\omega$ is simultaneously transferred to molecule D, which consequently transitions from a vibrational level $\nu_{\mathrm{D}}$ of the electronic ground state $\phi_{\mathrm{D}}$ to $\nu_{\mathrm{D}^+}$ of its ionized ground state $\phi_{\mathrm{D}^+}$, while emitting an electron of kinetic energy $\varepsilon'$.

Inserting Eq.~\eqref{eq:BH_expansion} into \eqref{eq:M} and employing the Born–Oppenheimer approximation, the transition matrix element within the first order of perturbation theory for a single initial and final state is
\begin{equation}
\label{eq:M_BO}
    M = 
    \langle \nu_{\mathrm{D}^+} | \langle \phi_{\mathrm{D}^+, \mathbf{k}'} |  \,
    \langle \nu_{\mathrm{A}^-} | \langle \phi_{\mathrm{A}^-} |
    \; \hat{V}_C \;
    | \phi_{\mathrm{A}, \mathbf{k}} \rangle | \nu_\mathrm{A} \rangle \, 
    | \phi_\mathrm{D} \rangle | \nu_\mathrm{D} \rangle
\end{equation}
where the momentum-normalized electronic scattering state $\phi_{\mathrm{A}, \mathbf{k}}$ ($\phi_{\mathrm{D}^+, \mathbf{k'}}$) describes A (D$^+$) together with the incoming (outgoing) electron.

For a sufficiently large distance $R_\mathrm{AD}$ between A and D, the Coulomb term $\hat{V}_C$ can be represented in a multipole expansion.
The first non-vanishing contribution to the transition matrix element $M$ corresponds to the dipole–dipole interaction of the two charge densities located on A and D,
\begin{equation}
    \label{eq:V}
    \hat{V}_{\mu\mu} 
    = \frac{1}{R_\mathrm{AD}^3} 
    \big[
    \hat{\boldsymbol{\mu}}_\mathrm{A} \cdot  \hat{\boldsymbol{\mu}}_\mathrm{D} - 3 \, (\mathbf{u}_R \cdot \hat{\boldsymbol{\mu}}_\mathrm{A}) \, (\mathbf{u}_R \cdot \hat{\boldsymbol{\mu}}_\mathrm{D})
    \big],
\end{equation}
where $\mathbf{u}_R$ is the unit vector pointing from A to the direction of D 
and $\hat{\boldsymbol{\mu}}_\mathrm{A}$ ($\hat{\boldsymbol{\mu}}_\mathrm{D}$) is the dipole moment operator of A (D), which includes all charged particles of A (D).\cite{gokhberg2009, cederbaum_ultrafast_2018}

Substituting Equations~\eqref{eq:M_BO} and \eqref{eq:V} into Eq.~\eqref{eq:diff_xs} yields the differential cross section in the lowest-order approximation to ICEC,
\begin{multline}
    \dv{\sigma}{\Omega_{\mathbf{k}'}} 
    = \frac{1}{(2\pi)^2}\frac{k'}{k}
    \frac{1}{R_\mathrm{AD}^3} 
    \big|
    \boldsymbol{\mu}_{\nu_\mathrm{A} \nu_{\mathrm{A}^-}} \cdot \boldsymbol{\mu}_{\nu_\mathrm{D} \nu_{\mathrm{D}^+}} \\ 
    - 3 \, (\mathbf{u}_R \cdot \boldsymbol{\mu}_{\nu_\mathrm{A} \nu_{\mathrm{A}^-}}) \, (\mathbf{u}_R \cdot \boldsymbol{\mu}_{\nu_\mathrm{D} \nu_{\mathrm{D}^+}}) 
    \big|^2,
\end{multline}
where the vibronic transition dipole elements are defined as
\begin{equation}
\begin{aligned}
    \label{eq:mu}
    \boldsymbol{\mu}_{\nu_\mathrm{A} \nu_{\mathrm{A}^-}} 
    &= \langle \nu_{\mathrm{A}^-} | \; 
    \langle \phi_{\mathrm{A}^-} | \; 
    \hat{\boldsymbol{\mu}}_\mathrm{A} \; 
    | \phi_{\mathrm{A}, \mathbf{k}} \rangle \; 
    | \nu_\mathrm{A} \big\rangle \\
    \boldsymbol{\mu}_{\nu_\mathrm{D} \nu_{\mathrm{D}^+}} 
    &= \langle \nu_{\mathrm{D}^+} | \; 
    \langle \phi_{\mathrm{D}^+, \mathbf{k'}} | \; 
    \hat{\boldsymbol{\mu}}_\mathrm{D} \; 
    | \phi_\mathrm{D} \rangle \; 
    | \nu_\mathrm{D} \rangle.
\end{aligned}
\end{equation}

We average over the orientation of A and D, which is equivalent to averaging over the transition dipole elements, leading to
\begin{equation}
    \frac{d\sigma}{d\Omega_{k'}} 
    = \frac{1}{(2\pi)^2} \, \frac{k'}{k} \, \frac{2}{3} \, \frac{|\boldsymbol{\mu}_{\nu_\mathrm{A} \nu_{\mathrm{A}^-}}|^2 \, |\boldsymbol{\mu}_{\nu_\mathrm{D} \nu_{\mathrm{D}^+}}|^2}{R_\mathrm{AD}^6}.
\end{equation}

The total cross section is obtained by averaging over the solid angle of the incoming electron $\Omega_k$ and integrating over the solid angle of the outgoing electron, $\Omega_{k'}$. 
For this, we expand the plane waves in the initial and final states in terms of spherical waves \cite{gokhberg2010, sobelman_1973}. 
We then multiply with $k k'$ to change from momentum to energy normalization of the continuum states, arriving at the total cross section
\begin{equation}
\label{eq:ICEC_explicit}
    \sigma
    = \frac{1}{(2\pi)^2}  \frac{(2\pi)^6}{4\pi k^2} \, \frac{2}{3} \, \frac{|\boldsymbol{\mu}_{\nu_\mathrm{A} \nu_{\mathrm{A}^-}}|^2 \, |\boldsymbol{\mu}_{\nu_\mathrm{D} \nu_{\mathrm{D}^+}}|^2}{R_\mathrm{AD}^6}.
\end{equation}

\begin{figure}[!ht]
    \centering
    \includegraphics[width=0.9\linewidth]{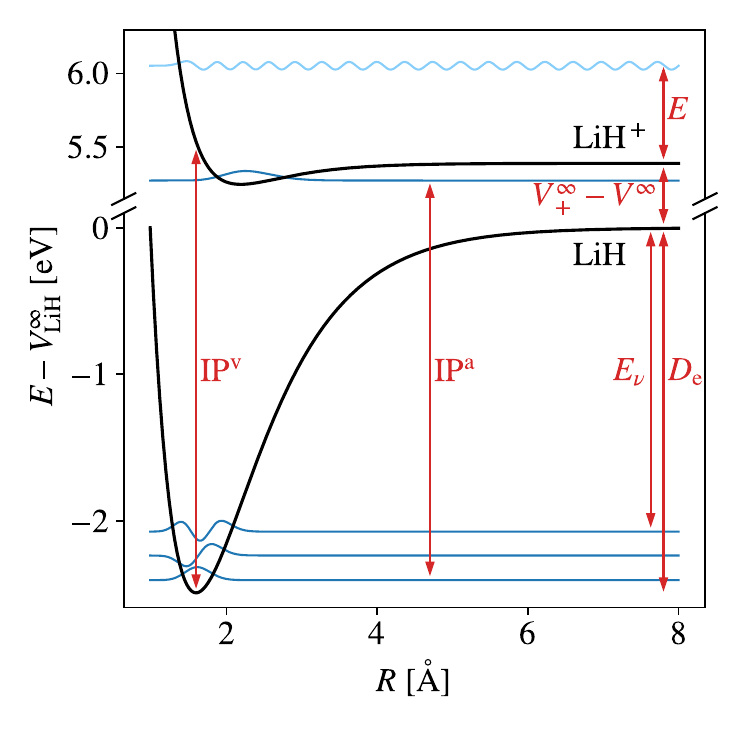}
    \caption{Potential energy surfaces (black) of the electronic ground states of LiH and LiH$^+$, including the lowest vibrational states and a dissociative state (blue).
    The dissociation energy $D_\mathrm{e}$, adiabatic and vertical ionization energies $\mathrm{IP}^\mathrm{a}$ and $\mathrm{IP}^\mathrm{v}$, bound and dissociative vibrational energies $E_\nu$ and $E$, respectively, are included for the readers' convenience.
    $V^\infty_+ - V^\infty$ is the difference between the surfaces at $R\to\infty$.
    Figure taken from~\onlinecite{figshare_figures}.}
    \label{fig:PES_LiH}
\end{figure}

The transition dipole elements defined in Eq.~\eqref{eq:mu} can be related to the corresponding vibrationally resolved photorecombination (PR) and photoionization (PI) cross sections. 
The PR cross section for the electron acceptor A is given by\cite{cederbaum_ultrafast_2018, sobelman_1973}
\begin{equation}
\label{eq:PR_XS}
    \sigma^\mathrm{PR}_{\nu_\mathrm{A} \nu_{\mathrm{A}^-}}(\varepsilon)
    =  \frac{(2\pi)^2 \omega^3}{3 c^3 k^2} \, 
    |\boldsymbol{\mu}_{\nu_\mathrm{A} \nu_{\mathrm{A}^-}}|^2,
\end{equation}
where $c$ is the speed of light and $\varepsilon$ is the kinetic energy of the incoming electron, corresponding to momentum $k$. 
The energy of the emitted photon, $\omega$, is the excess energy of the electron capture in the context of ICEC depicted in Fig.~\ref{fig:icec_sketch}.
Its value is given by the sum of $\varepsilon$ and the energy difference between the initial and final vibronic state
\begin{equation}
\label{eq:photon_energy}
   \omega = \epsilon + \big(V^\infty_\mathrm{A} + E_{\nu_\mathrm{A}}\big) - \big( V^\infty_{\mathrm{A^-}} + E_{\nu_{\mathrm{A}^-}} \big).
\end{equation}
Here, we define the energy of the $\nu$th vibrational state $E_\nu$ with respect to the $R\to\infty$ limit, $V^\infty$, of the corresponding potential energy surface (PES), as depicted in Fig.~\ref{fig:PES_LiH}.
This deviates from the conventional approach of defining with respect to the minimum of the potential energy curve but allows for a unified description of bound and dissociative vibrational states.
Bound states have then negative energies while dissociative (continuum) states are characterized by positive energies. 
For historical reasons, the equivalent equation 
$\omega 
= \epsilon + \mathrm{IP}^\text{\,a}_{\mathrm{A}^-} 
+ [(E_{\nu_\mathrm{A}}- E_{0_\mathrm{A}}) - (E_{\nu_{\mathrm{A}^-}}-E_{0_{\mathrm{A}^-}})]$
is implemented with the adiabatic ionization energy
$\mathrm{IP}^\text{\,a}_{\mathrm{A}^-} = (V^\infty_\mathrm{A} + E_{\nu_\mathrm{A} = 0}) - (V^\infty_{\mathrm{A}^-} + E_{\nu_{\mathrm{A}^-} = 0})$ as illustrated in Fig.~\ref{fig:PES_LiH}.

Similarly, the PI cross section for the electron donor D is\cite{lucchese_PI-XS, oneil_photoionization_1978}
\begin{equation}
\label{eq:PI_XS}
     \sigma^{\mathrm{PI}}_{\nu_\mathrm{D} \nu_{\mathrm{D}^+}}(\omega) 
    = \frac{ (2\pi)^2 \omega}{3 c} \,  
     |\boldsymbol{\mu}_{\nu_\mathrm{D} \nu_{\mathrm{D}^+}}|^2,
\end{equation}
where the outgoing electron has a kinetic energy of
\begin{equation}
\label{eq:final_kin_E}
    \varepsilon' = \omega 
    + \big(V^\infty_\mathrm{D} + E_{\nu_\mathrm{D}}\big)
    - \big(V^\infty_{\mathrm{D}^+} + E_{\nu_{\mathrm{D}^+}}\big)
\end{equation}
with the quantities analogously defined as in Eq.~\eqref{eq:photon_energy}.
As before, the equivalent equation
$\varepsilon' 
= \omega 
- \mathrm{IP}^\mathrm{\, a}_{\mathrm{D}} 
- [(E_{\nu_{\mathrm{D}^+}}- E_{0_{\mathrm{D}^+}}) - (E_{\nu_\mathrm{D}}-E_{0_\mathrm{D}})]$
is implemented for historical reasons.

With Equations~\eqref{eq:ICEC_explicit}, \eqref{eq:PR_XS} and \eqref{eq:PI_XS}, we arrive at an approximation to the cross section for ICEC between two interacting molecules
\begin{equation}
    \label{eq:intra_icec}
    \sigma(\varepsilon, \, \nu_\mathrm{A}\nu_\mathrm{D} \to \nu_{\mathrm{A}^-}\nu_{\mathrm{D}^+}) 
    = 
    \frac{3 c^4 }{4 \pi} \,
    \frac{
    \sigma^\mathrm{PR}_{\nu_\mathrm{A} \nu_{\mathrm{A}^-}}(\varepsilon) \; 
    \sigma^{\mathrm{PI}}_{\nu_\mathrm{D} \nu_{\mathrm{D}^+}}(\omega)
    }{\omega^4 \, R_\mathrm{AD}^6}
\end{equation}
for a single vibronic transition per molecule.
Transitions to dissociative vibrational states can be included by substituting with the corresponding dissociation cross sections.

The energy balance of this process follows from combining Eq.~\eqref{eq:photon_energy} and \eqref{eq:final_kin_E}:
\begin{multline}
\label{eq:energy_conservation}
    \varepsilon' 
    + \big( V^\infty_{\mathrm{A^-}} + E_{\nu_{\mathrm{A}^-}} \big)
    + \big(V^\infty_{\mathrm{D}^+} + E_{\nu_{\mathrm{D}^+}}\big)
    \\
    = \epsilon 
    + \big(V^\infty_\mathrm{A} + E_{\nu_\mathrm{A}}\big)
    + \big(V^\infty_\mathrm{D} + E_{\nu_\mathrm{D}}\big).
\end{multline}

\subsection{Transitions between multiple vibrational states}
\label{sec:theory_sum}

So far, we have considered a single vibrational transition per molecule. 
In practice, ICEC involves transitions between multiple vibrational states.
Assuming both molecules start in the ground vibrational state, we sum over bound and integrate over dissociative final vibrational states of energies $E_{\mathrm{A}^-}$ and $ E_{\mathrm{D}^+}$,
\begin{equation}
\label{eq:sum_final_states}
\begin{aligned}
    \sigma(&\varepsilon, 0\,0) \\
    = &\sum_{\nu_{\mathrm{A}^-}\, \nu_{\mathrm{D}^+}} \sigma(\varepsilon, 0\,0 \to \nu_{\mathrm{A}^-}\nu_{\mathrm{D}^+}) \\
    &+\sum_{\nu_\mathrm{A^-}} \int_{0}^{E_{\mathrm{D}^+}^\mathrm{max}} 
    \frac{
    \mathrm{d}\sigma(\varepsilon, 0\,0 \to \nu_{\mathrm{A}^-}E_{\mathrm{D}^+})
    }{
    \mathrm{d} E_{\mathrm{D}^+}
    }
    \, \mathrm{d} E_{\mathrm{D}^+} \\
    &+\sum_{\nu_{\mathrm{D}^+}} \int_{0}^{E_{\mathrm{A}^-}^\mathrm{max}} 
    \frac{
    \mathrm{d}\sigma(\varepsilon, 0\,0 \to E_{\mathrm{A}^-}\nu_{\mathrm{D}^+})
    }{
    \mathrm{d} E_{\mathrm{A}^-}
    }
    \, \mathrm{d} E_{\mathrm{A}^-} \\
    &+ \int_{0}^{E_{\mathrm{D}^+}^\mathrm{max}}  \int_{0}^{E_{\mathrm{A}^-}^\mathrm{max}} 
    \frac{
    \mathrm{d}\sigma(\varepsilon, 0\,0 \to E_{\mathrm{A}^-}E_{\mathrm{D}^+})
    }{
    \mathrm{d} E_{\mathrm{A}^-} \, \mathrm{d} E_{\mathrm{D}^+}
    }
    \, \mathrm{d} E_{\mathrm{A}^-} \, \mathrm{d} E_{\mathrm{D}^+}.
\end{aligned}
\end{equation}
For a given incoming energy $\varepsilon$ and final vibrational state $\nu_\mathrm{A^-}$ or $\nu_\mathrm{D^+}$, the maximum vibrational energy of $\mathrm{D}^+$ or $\mathrm{A}^-$, respectively, is given by
$\epsilon'= 0$ in Eq.~\ref{eq:energy_conservation} with $\nu_D = \nu_A =0$:
\begin{equation}
\label{eq:max_KER}
\begin{aligned}
    E^\text{max}_{\mathrm{D}^+} (\nu_{\mathrm{A}^-}) 
    = 
    \varepsilon 
    &+ \big[ V^\infty_\mathrm{A} + E_{\nu_\mathrm{A}=0}
    - V^\infty_{\mathrm{A^-}} - E_{\nu_{\mathrm{A}^-}}\big] \\
    &+ \big[V^\infty_\mathrm{D} + E_{\nu_\mathrm{D}=0} - V^\infty_{\mathrm{D}^+} \big] \\
    E^\text{max}_{\mathrm{A}^-} (\nu_{\mathrm{D}^+}) 
    = \varepsilon 
    &+ \big[V^\infty_\mathrm{A} + E_{\nu_\mathrm{A}=0} - V^\infty_{\mathrm{A^-}}\big] \\
    &+ \big[V^\infty_\mathrm{D} + E_{\nu_\mathrm{D}=0} - V^\infty_{\mathrm{D}^+} - E_{\nu_{\mathrm{D}^+}}\big]
\end{aligned}
\end{equation}
These values define the upper integration limits in Eq.~\eqref{eq:sum_final_states} to cover all possible distributions in energy between dissociation and the outgoing electron.
If energy conservation does not allow for dissociation of the molecule in question, the corresponding contribution is zero.

For dissociation of both molecules, described by the last term in Eq.~\eqref{eq:sum_final_states}, 
we need to cover all possible energy distributions between the outgoing electron and the dissociation of both $\mathrm{A}^-$ and $\mathrm{D}^+$.
We fix $E^\text{max}_{\mathrm{D}^+}$ by taking A at the minimum dissociation energy $E_{\mathrm{A}^-} = 0$, while $E^\text{max}_{\mathrm{A}^-}$ still depends on $E_{\mathrm{D}^+}$,
\begin{equation}
\label{eq:max_KER_diss}
\begin{aligned}
    E^\text{max}_{\mathrm{D}^+} 
    = \varepsilon &+ \big[V^\infty_\mathrm{A} + E_{\nu_\mathrm{A}=0} - V^\infty_{\mathrm{A^-}} \big] \\
    &+ \big[V^\infty_\mathrm{D} + E_{\nu_\mathrm{D}=0} - V^\infty_{\mathrm{D}^+} \big] \\
    E^\text{max}_{\mathrm{A}^-}
    = \varepsilon &+ \big[V^\infty_\mathrm{A} + E_{\nu_\mathrm{A}=0} - V^\infty_{\mathrm{A^-}} \big] \\
    &+ \big[V^\infty_\mathrm{D} + E_{\nu_\mathrm{D}=0} - V^\infty_{\mathrm{D}^+} - E_{\mathrm{D}^+}\big].
\end{aligned}
\end{equation}

Finally, to account for a thermal population in the initial states, we average over initial vibrational states using two Boltzmann distributions,
\begin{equation}
    \label{eq:boltzmann}
\begin{aligned}
    \sigma (\varepsilon, T) 
    = &\frac{1}{\sum_{\nu_\mathrm{A},\nu_\mathrm{D}} e^{-(E_{\nu_\mathrm{A}}+E_{\nu_\mathrm{D}})/(k_B T)}} \\ 
    & \times \sum_{\nu_\mathrm{A},\nu_\mathrm{D}}  e^{-(E_{\nu_\mathrm{A}}+E_{\nu_\mathrm{D}})/(k_B T)} \;
    \sigma (\varepsilon, \nu_{\mathrm{A}}\, \nu_{\mathrm{D}}),
\end{aligned}
\end{equation}
where $\sigma (\varepsilon, \nu_{\mathrm{A}}\, \nu_{\mathrm{D}})$ is defined analogous to Eq.~\eqref{eq:sum_final_states}, $T$ is the thermodynamic temperature, and $k_B$ is the Boltzmann constant.
This temperature dependence is a key feature of our model, made possible by explicitly including nuclear motion.

\subsection{Asymptotic ICEC model with the Franck-Condon principle}
\label{sec:theory_FC}

Ideally, experimental or accurate numerical vibrationally resolved PI cross sections are available for all molecules we want to study. 
This is, however, only the case for a limited number of systems.
The Condon approximation\cite{Herzberg1950} remedies this for diatomic molecules, by assuming that the electronic transition dipole element $\boldsymbol{\mu}$ is independent of $R$, 
\begin{equation}
    \boldsymbol{\mu} (R, \epsilon) 
    = \langle \phi_{\mathrm{D}^+, \mathbf{k'}} | \,\hat{\boldsymbol{\mu}} \, | \phi_\mathrm{D} \rangle 
    \approx \boldsymbol{\mu} (R_e, \epsilon),
\end{equation}
which yields an approximation to the PI cross section in Eq.~\eqref{eq:PI_XS} of the form
\begin{equation}
\label{eq:PI_xs_FC}
    \sigma^{\mathrm{PI}}_{\nu_\mathrm{D}\nu_{\mathrm{D}^+}}(\omega) \approx \frac{4\pi^2 \omega}{3 c} |\mu(R_e, \epsilon)|^2 |\langle \nu_\mathrm{D} | \nu_{\mathrm{D}^+} \rangle |^2
\end{equation}
including the Franck-Condon factors $|\langle \nu_\mathrm{D} | \nu_{\mathrm{D}^+} \rangle |^2 $.
This approximation is valid if  
the cross section is dominated by the vertical transition at the equilibrium distance $R_e$ of the initial PES
and when the transition is free of autoionizing states and shape resonances.

This leaves the challenge of calculating the transition dipole moments explicitly.
Let us have a look at the energies at play.
From energy conservation in Eq.~\eqref{eq:final_kin_E}, we know that
$\omega + V^\infty_\mathrm{D} + E_{\nu_\mathrm{D}} = \varepsilon' + V^\infty_{\mathrm{D}^+} + E_{\nu_{\mathrm{D}^+}}$ 
whereas for electronic vertical transitions we have
$\omega + V_\mathrm{D}(R_e) = \bar{\varepsilon}' + V_{\mathrm{D}^+}(R_e)$.
Approximating the transition dipole moment with its purely electronic counterpart, $\mu(R_e, \epsilon') \approx \mu(R_e, \bar{\epsilon}')$, we can identify the unresolved cross section to be
\begin{equation}
    \sigma^{\mathrm{PI}}(\omega) \approx \frac{4\pi^2 \omega}{3 c} |\mu(R_e, \bar{\epsilon}')|^2
\end{equation}
giving an approximation for the vibrationally resolved cross section
\begin{equation}
\label{eq:PI_xs_energy}
    \sigma^{\mathrm{PI}}_{\nu_\mathrm{D}\nu_{\mathrm{D}^+}}(\omega) \approx \sigma^{\mathrm{PI}}_\mathrm{D}(\omega) \, |\langle \nu_\mathrm{D} | \nu_{\mathrm{D}^+} \rangle |^2. 
\end{equation}
Here, $\bar{\epsilon}'$ does not depend on $\nu_\mathrm{D}$ and $\nu_{\mathrm{D}^+}$ anymore, introducing a violation in energy conservation compared to Eq.~\eqref{eq:final_kin_E} and making this approach best suited for small changes in vibrational energy.

Applying this to ICEC in Eq.~\ref{eq:ICEC_explicit}, or directly in Eq.~\ref{eq:intra_icec}, the Franck-Condon based ICEC cross section is given by
\begin{equation}
    \label{eq:intra_icec_FC}
    \begin{aligned}
    \sigma(\varepsilon, \, \nu_\mathrm{A}\nu_\mathrm{D} \to \nu_{\mathrm{A}^-}\nu_{\mathrm{D}^+}) 
    &= 
    \frac{3 c^4 }{4 \pi} \,
    \frac{
    \sigma^\mathrm{PR}_\mathrm{A}(\varepsilon) \; 
    \sigma^\mathrm{PI}_\mathrm{D}(\omega)
    }{\omega^4 \, R_\mathrm{AD}^6} \\ 
    &\quad \times |\langle \nu_{\mathrm{A}^-} | \nu_{\mathrm{A}} \rangle |^2 
    \, |\langle \nu_\mathrm{D} | \nu_{\mathrm{D}^+} \rangle |^2  
    \end{aligned}
\end{equation}
with $\varepsilon$ and $\omega$ according to Eq.~\ref{eq:photon_energy}.

We note that in the limit $\varepsilon\to\infty$ -- where all final vibrational states are energetically accessible -- the sum (and integration) over all final bound and dissociative states of D, $\nu_{\mathrm{D}^+}$ and $E_{\mathrm{D}^+}$, in Eq.~\eqref{eq:sum_final_states}
removes the dependency of the ICEC cross section on the nuclear dynamics of D.
The complete sum over the Franck-Condon factors for a given initial vibrational state can be identified in Eq.~\eqref{eq:sum_final_states} and is equal to one.
This is not possible for the sum over $\nu_{\mathrm{A}^-}$ and $E_{\mathrm{A}^-}$ in Eq.~\eqref{eq:sum_final_states}, as $\omega$ in Eq.~\ref{eq:photon_energy}, and thus $ \sigma^\mathrm{PI}_\mathrm{D}(\omega)$, depend on the vibronic transition of A.
For finite $\varepsilon$, the summation in Eq.~\eqref{eq:sum_final_states} only runs over the energetically accessible final states, affecting the completeness of the sum over Franck-Condon factors.

The Franck-Condon principle can be extended to polyatomic molecules\cite{FC_molecules_1964, h2o_2019} where vibrational states are defined by a set of quantum numbers $\{\nu\}$.

\subsection{Vibrational branching ratios for molecular cross sections}
\label{sec:theory_BR}

Another option to obtain the vibrationally resolved PI cross sections are the branching ratios, provided they are available together with the partial cross section $\sigma^\mathrm{PI}_\nu = \sum_{\nu_+} \sigma^\mathrm{PI}_{\nu \nu_+}$.
Unfortunately, these are also not often reported in literature.

One definition for the branching ratio is the cross section corresponding to one vibrational level divided by the sum $\sigma^\mathrm{PI}_\nu$
\begin{equation}
    \mathrm{BR}_{\nu\nu_+} (\omega) 
    = \frac{\sigma^{\mathrm{PI}}_{\nu \nu_+} }{\sigma^\mathrm{PI}_\nu}
    .
\end{equation}
If $\sigma_\nu$ does not exist in literature, we may approximate it by the unresolved cross section $\sigma^\mathrm{PI}(\omega)$,
\begin{equation}
    \sigma^{\mathrm{PI}}_{\nu\nu_+} (\omega) 
    = \mathrm{BR}_{\nu\nu_+} (\omega) \,
    \sigma^\mathrm{PI} (\omega).
\end{equation}

Another definition is the ratio between cross sections of different final vibrational states (also called $v$-ratios):
\begin{equation}
    \frac{\sigma^{\mathrm{PI}}_{\nu\nu_+} }{
    \sigma^{\mathrm{PI}}_{\nu \nu_+'} 
    }.
\end{equation}
Together with $\sigma^\mathrm{PI}_\nu$ they give rise to a set of linear equations at each energy which can be solved to get $\sigma^\mathrm{PI}_{\nu \nu_+}$.
Within the Franck-Condon principle, $v$-ratios are equal to the ratio of the respective Franck-Condon factors\cite{h2o_2019}.

With our theoretical foundations in place, let us introduce our physical system we shall investigate.

\section{System studied: \ce{H+} LiH}
\label{sec:systems}

Both Lithium hydride (\ce{LiH}) and protons (\ce{H+}) occurred in the early universe alongside free electrons and their processes are central to cosmochemistry\cite{Li_2003, HLiH_2023}.
LiH is one of the few molecules with reported vibrationally resolved PI cross sections for multiple initial vibrational states\cite{lundsgaard_LiH}, though these theoretical data only include bound-bound transitions. 
Nevertheless, these cross sections enable benchmarking the Franck-Condon model given in Eq.~\eqref{eq:intra_icec_FC}, which can be extended to include dissociation.
To the best of our knowledge, vibrational branching ratios for LiH are not reported in the literature.

Assuming the electron acceptor to be an atom 
removes the sum and integration over the final states $\nu_\mathrm{A^-}$ and $E_\mathrm{A^-}$ in Eq.~\eqref{eq:sum_final_states}.

\ce{H+} and \ce{LiH} are further ideal candidates due to their PR and PI cross sections being free of resonance or autoionization features, allowing for a clearer analysis in this initial study.

Vibrational energies of LiH and LiH$^+$ are obtained from Morse potentials with the limit $V(R\to\infty) = 0$ to comply with the definition of vibrational energies, $E_\nu < 0$, in Sec.~\ref{sec:theory_asymptotic},
\begin{equation}
    \label{eq:Morse}
    V(R) = D_\mathrm{e} \left( 1 - e^{- \omega_\mathrm{e}\sqrt{\frac{\mu}{2 D_\mathrm{e}}} (R-R_\mathrm{e})^2} \right) - D_\mathrm{e},
\end{equation}
fitted to the ground electronic states of LiH and LiH$^+$ and defined by the parameters in Tab.~\ref{tab:parameter}.
The corresponding potential energy surfaces (PES) shown in Fig.~\ref{fig:PES_LiH} support 30 and 5 bound states, while \textit{ab initio} studies result in 24 and 7 states\cite{lundsgaard_LiH}, respectively.  
This deviation arises from the Morse potential being an approximation to the \textit{ab initio} PES of LiH and \ce{LiH+}\cite{lundsgaard_LiH}.

\begin{table*}[ht]
    \centering
    \caption{
    Relevant properties of H and LiH:
    ionization potential IP, photoionization cross section $\sigma_\mathrm{PI}$ at $\SI{14.6}{eV}$, reduced mass $\mu$, equilibrium distance $R_\text{e}$, dissociation energy $D_\mathrm{e}=\omega_\mathrm{e}^2/(4\omega_\mathrm{e}x_\mathrm{e})$, and harmonic vibrational frequency $\omega_\mathrm{e}$.}
    \begin{tabular}{l@{\hskip 8pt}c@{\hskip 10pt}c@{\hskip 10pt}c@{\hskip 10pt}c@{\hskip 10pt}c@{\hskip 10pt}c}
        \toprule
        & $\mathrm{IP}\, [\unit{eV}]$ 
        & \;$\sigma_\mathrm{PI}\, [\unit{Mb}]$
        & $\mu \, [\unit{a.u.}]$ 
        & $R_\text{e} \, [\unit{a.u.}]$
        & $D_\mathrm{e} \, [\unit{eV}]$
        & $\omega_\text{e} \, [\unit{cm^{-1}}]$
        \\ \midrule
        H
        & 13.6\cite{NIST}\;\,\,
        & \;5.23\cite{topbase1992}
        \\
        LiH ($\mathrm{X}\,^1\Sigma^+$)
        & \;\;\;7.7\cite{huber_molSpectra, lundsgaard_LiH}
        & \;7.13\cite{lundsgaard_LiH}
        & \;1618.09\cite{huber_molSpectra} 
        & 3.0148\cite{lundsgaard_LiH} 
        & 2.4924\cite{lundsgaard_LiH}\;\:
        & 1406.18\cite{lundsgaard_LiH}
        \\
        LiH$^+$ ($\mathrm{X}\,^2\Sigma^+$)
        & & &
        & 4.136\cite{lundsgaard_LiH}\;\:
        & 0.14374\cite{lundsgaard_LiH} 
        & 442.9\cite{lundsgaard_LiH}
        \\
        \bottomrule
    \end{tabular}
    \label{tab:parameter}
\end{table*}

The ionization energies of LiH and H, equal to the electron affinity of \ce{H+}, are listed in Tab.~\ref{tab:parameter}.
The ionization potential of LiH\cite{lundsgaard_LiH} is defined as the energy difference at the equilibrium values of the LiH and the \ce{LiH+} PES in Fig.~\ref{fig:PES_LiH}.
The adiabatic ionization energy is offset by the ground vibrational energies, $\mathrm{IP}^\text{\,a}_\mathrm{LiH} = \mathrm{IP}_\mathrm{LiH} + (D^{\mathrm{LiH}^+}_\mathrm{e} + E_{\nu_{\mathrm{LiH}^+}=0}) - (D^\mathrm{LiH}_\mathrm{e} + E_{\nu_\mathrm{LiH}=0})$.
Energy conservation implies that when $\mathrm{H}^+$ captures an electron, the electron affinity is sufficient to ionize LiH.

The asymptotic model in Eq.~\ref{eq:intra_icec} forms a lower bound to the cross section of ICEC, increasing in accuracy with larger distances $R_\mathrm{AD}$ between the center of masses of A and D \cite{bande_2023}.
Ref.~\cite{HLiH_2023} reports a minimum in the potential energy for the colinear geometry of $[\mathrm{Li H H}]^+$ at $R_\mathrm{LiH}=\SI{1.65}{\angstrom}$ and $R_{\mathrm{HH}^+}=\SI{2.51}{\angstrom}$ which corresponds to LiH weakly interacting with a proton.
We use the resulting distance between the center of mass of LiH and \ce{H+} of $R_\mathrm{AD} = \SI{3.95}{\angstrom}$ for our model.

\section{Computational Details}
\label{sec:comp-details}

The ICEC model requires PR and PI cross sections of \ce{H+} and LiH, respectively, and vibrational states of LiH. 
All calculations were performed with mol-ICEC\cite{zenodo_mol-ICEC}, which is based on the Python module mpmath\cite{mpmath}.

The PR cross section of \ce{H+} is obtained with the principle of detailed balance \cite{sobelman_1973} from the PI cross section:
\begin{equation}
    \sigma^\mathrm{PR}_\mathrm{A}(\varepsilon)
    = \frac{\omega^2}{2\varepsilon c^2}\, \frac{g_{\mathrm{A}^-}}{g_\mathrm{A}} \, \sigma^\mathrm{PI}_{\mathrm{A}^-}(\omega)
\end{equation}
where $\omega$ and $\varepsilon$ are connected by energy conservation in Eq.~\eqref{eq:photon_energy} and $g_{\mathrm{A}}$ and $g_{\mathrm{A}^-}$ are the multiplicities of the corresponding states.
For H, we have $g_\mathrm{H}/g_{\mathrm{H}^+} = 2 $

For LiH\cite{lundsgaard_LiH}, the PI data was digitized with the WebPlotDigitizer\cite{WebPlotDigitizer}.
The electronic PI cross section corresponds to the vertical ionization at the equilibrium distance $R_\mathrm{e}$ of the electronic ground state of LiH\cite{lundsgaard_LiH}.
Unfortunately, only vibrationally resolved PI cross sections of LiH from the three lowest vibrational states of the electronic ground state to all bound vibrational states of the ionized ground state are reported.
For the \textit{ab initio} based model in Eq.~\ref{eq:intra_icec}, we use the vibrational energies calculated from the vibrational energy differences\cite{lundsgaard_LiH}.
The Franck-Condon based model discussed in Sec.~\ref{sec:theory_FC} uses the analytical solutions\cite{matsumoto_1988} to the Morse potential to calculate vibrational energies and Franck-Condon factors. 
Dissociative states are discretized and normalized within a finite box, $L \in (0, \SI{8}{\angstrom}]$.
Multiplying the corresponding cross sections with the density of states, $\rho(E_f) = 1/|E_{f+1} - E_f|$, provides energy-normalized cross sections.

\section{Results for \ce{H+} LiH}
\label{sec:results}

In this section, we present the results for \ce{H+} LiH, discussing the effect of vibrational motion of LiH on the total cross section, the electron spectrum, and the temperature dependence of ICEC.
Unless stated otherwise, we assume LiH starts in the lowest vibrational state of its ground electronic state and sum over all vibrational states of its ionized ground electronic state.
We denote the vibrational states of LiH and \ce{LiH+} by $\nu$ and $\nu_+$, respectively.

\subsection{Total ICEC cross section}
\label{sec:results_benchmark}

The objective is to include both bound and dissociative intraparticle nuclear dynamics into the ICEC model.
However, even for well-studied molecules like LiH, to the best of our knowledge, the literature reports only bound-bound photoionization (PI) cross sections \cite{lundsgaard_LiH}.
The Franck-Condon principle offers a straightforward framework to include the dissociation in ICEC.
We aim to validate the Franck-Condon approach to ICEC given in Sec.~\ref{sec:theory_FC}, ensuring its applicability to a broader range of molecules in future studies.

\subsubsection{Bound-bound transitions from Frack-Condon v.s. \textit{ab initio} photoionization cross sections}

\begin{figure}[!ht]
    \centering
    \includegraphics[width=\linewidth]{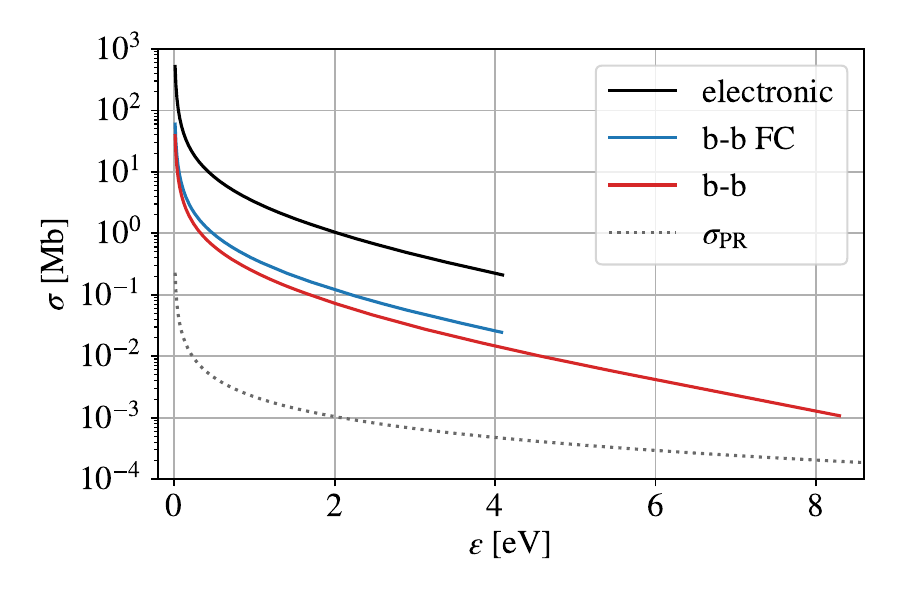}
    \caption{ICEC cross section for \ce{H+}-LiH vs. incoming electron energy $\varepsilon$ for $R_{\mathrm{H}\mathrm{-LiH}}=\SI{3.95}{\angstrom}$.
    Black: electronic results with vertical ionization of LiH at $R_\mathrm{e}$;
    red (b-b): with vibrationally resolved cross sections of LiH; 
    blue (b-b FC): based on Franck-Condon model.
    Only bound-bound transitions of LiH are included. 
    The photorecombination cross section of \ce{H+} (dotted gray) is included for comparison.
    Figures taken from Ref.~\onlinecite{figshare_figures}.}
    \label{fig:tot_xs}
\end{figure}

Figure~\ref{fig:tot_xs} shows total ICEC cross sections\cite{figshare_data} at a separation $R_\mathrm{AD} = \SI{3.95}{\angstrom}$ between the center of mass of \ce{H+} and LiH for three levels of approximation:
purely electronic ICEC (black),
the model given in Eq.~\eqref{eq:intra_icec} and \eqref{eq:sum_final_states} with vibrationally resolved \textit{ab initio} PI cross sections for bound-bound transitions of LiH (red, b-b),
and the Franck-Condon based model (Sec.~\ref{sec:theory_FC}) with Franck-Condon factors describing the bound-bound transitions (blue, b-b FC).
The photorecombination (PR) cross section of \ce{H+} (dotted gray) is included for comparison.

The electronic and the Franck-Condon results are available only up to kinetic energies of about $\varepsilon=\SI{4}{eV}$ because Lundsgaard \textit{et al.}~\cite{lundsgaard_LiH} report a narrower photon energy range for electronic than for vibrationally resolved PI cross sections of LiH.

All three results show a $\varepsilon^{-4}$ dependency of the ICEC cross section.
The kinetic energy $\varepsilon$ is connected to the transferred energy $\omega$ by energy conservation in Eq.~\eqref{eq:photon_energy} and the corresponding $\omega^{-4}$ term is present in the ICEC cross sections in Equations~\eqref{eq:icec_electronic}, \eqref{eq:intra_icec}, and \eqref{eq:intra_icec_FC}. 
Additional resonance pathways, which are absent in LiH and \ce{H+}, would introduce an overlaying structure to this dependency.

ICEC dominates over the PR of \ce{H+} at $R_\mathrm{AD} = \SI{3.95}{\angstrom}$ for small incoming electron energies,
enhancing the electron attachment compared to isolated \ce{H+}.
While both ICEC cross sections including nuclear motion still dominate over PR, they are one order of magnitude smaller than the ICEC cross section of the purely electronic simulation.

The ICEC cross section for \ce{H+} LiH starts at zero electron energy because the attachment energy of \ce{H+} is already sufficient to ionize LiH. 
If it were not, the threshold would be above zero\cite{gokhberg2010} and depend on vibrational states according to Eq.~\eqref{eq:final_kin_E} with $\varepsilon'=0$.
The threshold is decreased if the energy of the ground vibrational state of the neutral electron donor D is larger than the one of its cationic state $\mathrm{D}^+$.
Vibrational effects on thresholds were previously observed when including vibrational motion between ICEC partners\cite{jahr2025,fede2024}.

While the Franck-Condon results successfully capture the overall trend of the ICEC cross sections determined from theoretical \textit{ab initio} LiH PI cross sections, they deviate by a factor of 1.5 to 1.8 with increasing electron energies.
The approximations necessary for the Franck-Condon model in Eq.~\eqref{eq:intra_icec_FC} introduce an error to the cross section compared to employing more accurate \textit{ab initio} PI cross sections in Eq.~\eqref{eq:intra_icec}. 
This deviation is strongly system dependent and no general rule regarding an over- or underestimation can be reported.
The sum of our calculated Franck-Condon factors lies within 5\% of the sum of the factors from the \textit{ab initio} calculations in Ref.~\onlinecite{lundsgaard_LiH}.

Although ICEC results from vibrationally resolved \textit{ab initio} PI cross sections are assumed to be more accurate, we cannot include the dissociation of LiH unless the corresponding cross sections are available in the literature.
We have therefore only considered bound-bound transitions of LiH in this part of the manuscript.
The overlap between the bound vibrational states of the ground electronic states of LiH and \ce{LiH+}, however, is small, because the corresponding PES are shifted relative to each other (see Fig.~\ref{fig:PES_LiH}).
The vertical transition would thus favor dissociation of LiH, which is neglected in the results presented in Fig.~\ref{fig:tot_xs}.
The Franck-Condon ICEC model remedies this, offering a more complete description of ICEC in the case of \ce{H+} LiH.

\subsubsection{Including dissociation of LiH with the Franck-Condon principle}
\label{sec:results_FC-model}

\begin{figure}[!ht]
    \centering
    \includegraphics[width=\linewidth]{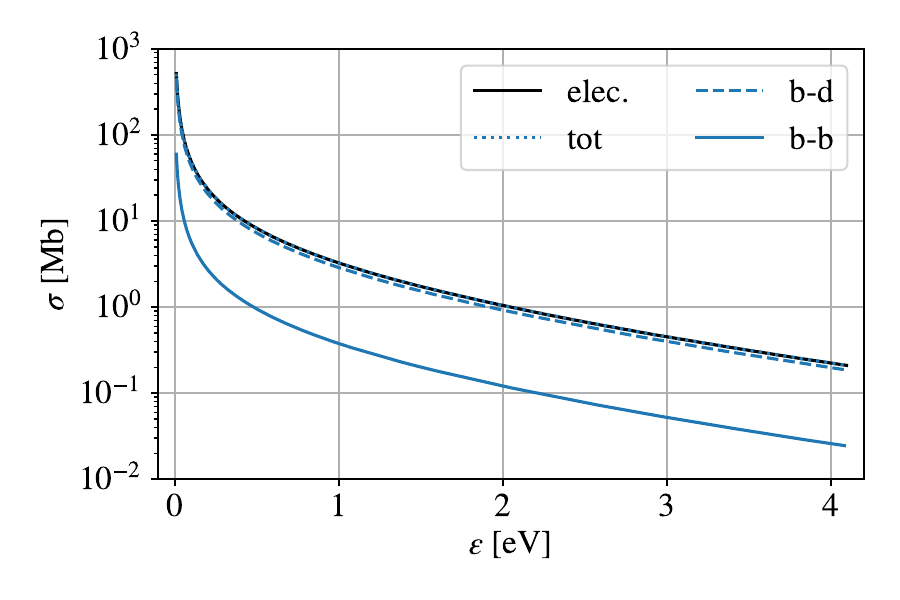}
    \caption{ICEC cross section for \ce{H+}-LiH vs. incoming electron energy $\varepsilon$.
    Black (elec.): electronic results from Fig.~\ref{fig:tot_xs}.
    Blue: includes nuclear dynamics of LiH within the Condon approximation; 
    solid (b-b): bound-bound transitions of LiH from Fig.~\ref{fig:tot_xs}, 
    dashed (b-d): bound-dissociative transitions, 
    dotted (tot): all transitions, coincides with electronic case.
    Figure taken from Ref.~\onlinecite{figshare_figures}.
    }
    \label{fig:tot_xs-FC}
\end{figure}

Figure~\ref{fig:tot_xs-FC} presents ICEC cross sections\cite{figshare_data} for \ce{H+} LiH from the Franck-Condon based model including both bound-bound (solid blue) and bound-dissociative transitions (dashed blue) of LiH.
The total cross section (dotted blue) is the sum of the bound and dissociative contributions.
The purely electronic ICEC cross section (black) is shown for comparison.

The cross section from bound-dissociative transitions of \ce{LiH} dominates over bound-bound transitions by one order of magnitude, indicating that LiH predominantly dissociates during ICEC.

The total cross section for one initial vibrational state now aligns with the electronic cross section.
In the case of \ce{H+} LiH, 
zero incoming electron energy already results in a transferred energy $\omega$ of $\SI{5.9}{eV}$ larger than needed for the ionization of LiH.
As \ce{LiH+} has a dissociation energy of $\SI{0.14}{eV}$ (see Tab.~\ref{tab:parameter}), all final dissociative states with energy below $\SI{5.76}{eV}$ are always accessible.
These states suffice to give approximately unity when their Franck-Condon factors are integrated and added to the sum over the corresponding bound-bound factors (see Sec.~\ref{sec:theory_FC}).
For the sum over all energetically accessible channels in Eq.~\eqref{eq:sum_final_states}, the electronic cross section in Eq.~\eqref{eq:icec_electronic} is therefore recovered to $>\SI{99.9}{\%}$.

These results highlight that including dissociative nuclear degrees of freedom in the model can be essential to accurately capture the nuclear dynamics in ICEC.

\subsection{ICEC electron spectrum}

We first benchmark the Franck-Condon model against the results\cite{figshare_data} from computational \textit{ab initio} PI cross sections for the available bound-bound transitions of the electron donor LiH and then show the spectrum including dissociation of LiH.

\subsubsection{Bound-bound transitions from Frack-Condon v.s. \textit{ab initio} photoionization cross sections}
\label{sec:results_benchmark_spectrum}
\begin{figure}[!ht]
    \centering
    \includegraphics[width=\linewidth]{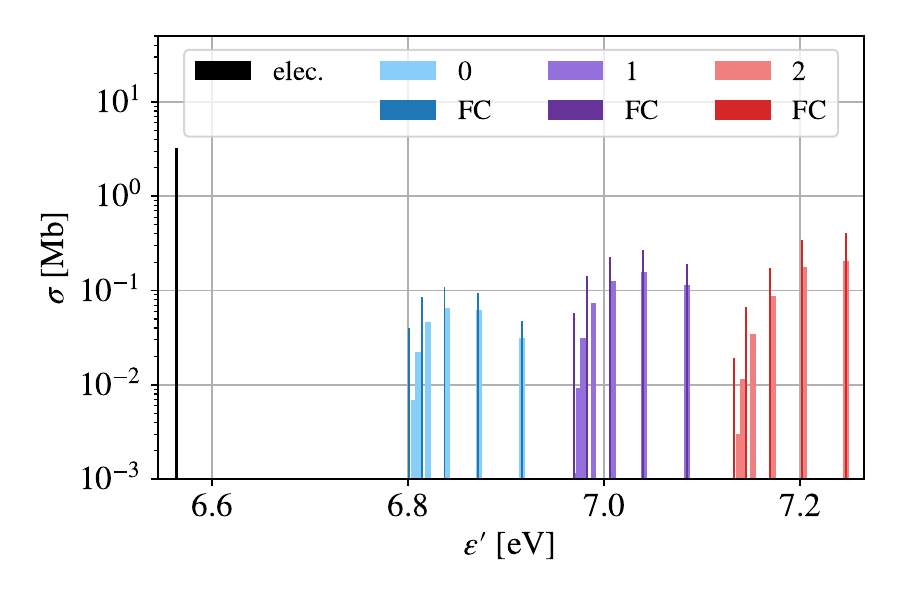}
    \caption{ICEC cross section for \ce{H+} LiH at $\varepsilon=\SI{1}{eV}$ and $R_{\mathrm{H}^+\mathrm{-LiH}}=\SI{3.95}{\angstrom}$ vs. outgoing electron energy $\varepsilon'$ for different initial vibrational states $\nu$ of LiH (blue: $\nu=0$, purple: $\nu=1$, red: $\nu=2$).
    Lighter shades (wider lines) include vibrationally resolved photoionization cross sections from theory, darker shades use the Franck-Condon principle.
    Each line corresponds to one transition from $\nu$ of LiH to $\nu_+$ of $\mathrm{LiH}^+$.
    Electronic case (black) corresponds to the vertical ionization of LiH. 
    Figure taken from Ref.~\onlinecite{figshare_figures}}
    \label{fig:spectrum_bb}
\end{figure}

Figure~\ref{fig:spectrum_bb} presents outgoing electron energy spectra\cite{figshare_data} for ICEC at a separation $R_\mathrm{AD} = \SI{3.95}{\angstrom}$ between the center of mass of \ce{H+} and LiH  with a fixed incoming kinetic energy of $\varepsilon = \SI{1}{\electronvolt}$
for three different vibrational initial states ($\nu=0,1,2$) and
for the three levels of approximation: purely electronic (black peak, Eq.~\eqref{eq:icec_electronic}), with vibrationally resolved PI cross sections (lighter, wider peaks,  Eq.~\eqref{eq:intra_icec}, and the Franck-Condon model (darker, narrower peaks, Eq.~\eqref{eq:intra_icec_FC}).
The spectrum for each vibrational initial state contains multiple peaks arising from different final vibrational states $\nu_+$ of \ce{LiH+}.

The peaks originating from higher vibrational initial states $\nu$ are located at higher kinetic energies of the emitted electron, while higher final vibrational states $\nu_+$ cause peaks at lower energies within their respective spectrum.
Higher $\nu$ reduces the required ionization energy of LiH (see also Fig.~\ref{fig:icec_sketch}), and therefore increases the kinetic energy of the outgoing electron $\varepsilon'$ for a fixed kinetic energy of the incoming electron.
For a specific initial state, more energy is required to reach higher vibrational final states of \ce{LiH+} rather than the vibrational ground state. 
Due to energy conservation, the corresponding peaks are found at lower kinetic energies of the emitted electron $\varepsilon'$.

We further see an increase of the highest individual cross section for the different vibrational initial states from $\nu=0$ (blue) to $\nu=2$ (red).
This can be understood by analyzing the overlap between the vibrational states $\nu$ and $\nu_+$ of the PES shown in Fig.~\ref{fig:PES_LiH}. 
For higher $\nu$, the nuclear wavefunctions extend to larger interatomic distances and thereby increase the overlap with the final nuclear wavefunction, consequently elevating the ICEC cross sections.

The Franck-Condon results (narrow lines) show fewer final states than the model based on the vibrationally resolved \textit{ab initio} PI cross sections. 
As discussed in Sec.~\ref{sec:systems},
the Morse potential approximating the \textit{ab initio} PES\cite{lundsgaard_LiH} supports fewer final states in the case of \ce{LiH+} and therefore, the number of peaks in the ICEC spectrum differs accordingly.
The corresponding peaks are increased by up to a factor of two, as discussed for the total cross section in Sec.~\ref{sec:results_benchmark}, but agree in overall trend.

The electronic cross section gives a single black line at $\varepsilon' = \varepsilon + \mathrm{IP}_\mathrm{H} - \mathrm{IP}^\mathrm{v}_\mathrm{LiH}$.
The energy is lower compared to vibrational results because it corresponds to a vertical transition of LiH at $R_e$ with vertical ionization energy $\mathrm{IP}^\mathrm{v}_\mathrm{LiH}$ between PES in Fig.~\ref{fig:PES_LiH} requiring more energy compared to bound vibronic transitions.
The electronic approximation fails to explain the position of the maximum and the broadening in the electron spectrum.

Now that we have established qualitative agreement in the electron spectrum between the Franck-Condon ICEC model and the results from \textit{ab initio} PI cross section, we shall include the dissociation of LiH.

\subsubsection{Including dissociation of LiH with the Franck-Condon principle}

\begin{figure}[!ht]
    \centering
    \includegraphics[width=\linewidth]{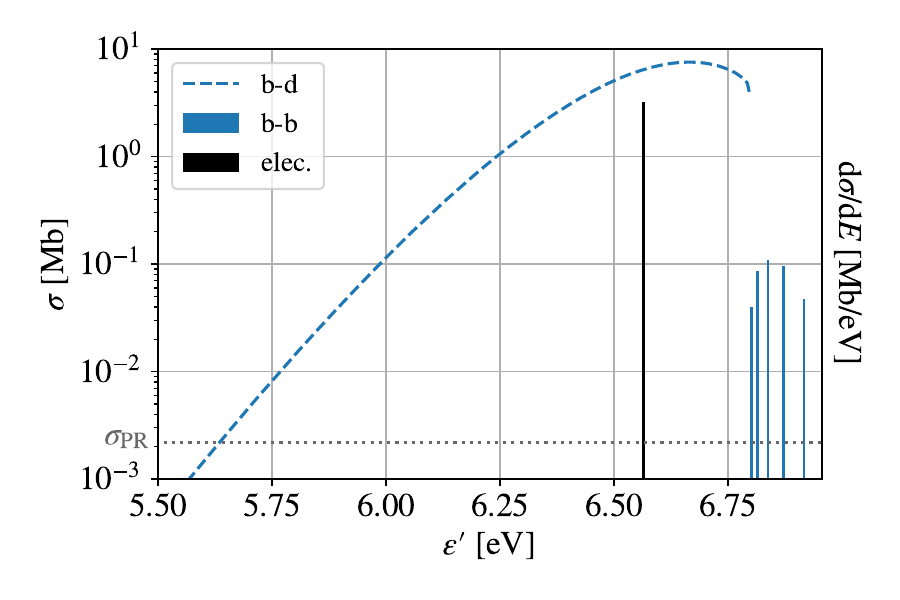}
    \caption{ICEC cross section for \ce{H+} LiH at $\varepsilon=\SI{1}{eV}$ vs. kinetic energy of the outgoing electron $\varepsilon'$ for the ground vibrational state of LiH determined from the Franck-Condon model (Eq.~\eqref{eq:intra_icec_FC}).
    Blue peaks involve bound-bound transitions during ionization of LiH;
    Blue dashed line (FC b-d) is the differential $\mathrm{d}\sigma/\mathrm{d}E$ involving bound-dissociative transitions.
    Electronic peak (black) from Fig.~\ref{fig:spectrum_bb}.
    Grey dotted: photorecombination cross section of \ce{H+} at $\varepsilon=\SI{1}{eV}$.
    Figure taken from Ref.~\onlinecite{figshare_figures}.}
    \label{fig:spectrum-FC}
\end{figure}

We present the ICEC electron spectrum\cite{figshare_data} for \ce{H+} LiH in Fig.~\ref{fig:spectrum-FC} including dissociation of LiH within the Condon approximation at a fixed incoming electron energy of $\epsilon = \SI{1}{eV}$.
LiH is initially in the ground vibrational state, $\nu = 0$. 
The spectrum comprises two distinct contributions: bound-bound transitions (solid blue, b-b) and bound-dissociative transitions (dashed blue curve, b-d).
The purely electronic ICEC cross section (black) and the fixed value of the PR cross section of \ce{H+} at $\varepsilon=\SI{1}{eV}$ (horizontal dotted gray) are included for comparison.

The inclusion of the dissociation of LiH introduces a bound-dissociative continuum represented by the differential cross section $\mathrm{d}\sigma/\mathrm{d}E$ in addition to the discrete bound-bound peaks observed in Fig.~\ref{fig:spectrum_bb}. 
Values at lower outgoing electron energies $\varepsilon'$ correspond to larger dissociation energies of LiH.
At $\epsilon = \SI{1}{eV}$, both the contributions from bound-bound transitions and dissociation up to $\SI{0.9}{eV}$ exceed the PR of $\ce{H+}$, as highlighted in Fig.\ref{fig:tot_xs-FC}.

Dissociation of LiH dominates the spectrum, consistent with the findings in the total ICEC cross section in Fig.~\ref{fig:tot_xs-FC}.
The spectrum peaks at $\varepsilon'=\SI{6.67}{eV}$ with an electron energy gain of $\varepsilon'- \varepsilon = \SI{5.67}{eV}$.
At this maximum, the dissociation energy of \ce{LiH+} is $\SI{0.13}{eV}$, indicating that ICEC predominantly leads to slowly separating nuclei of LiH.

As seen in Fig.~\ref{fig:spectrum_bb}, a purely electronic transition would produce a single peak, uniquely determined by the incoming energy and positioned at an outgoing energy of $\SI{0.1}{eV}$ below the maximum of the bound-dissociative spectrum.
This discrepancy indicates that neither the electronic model, nor bound-bound transitions alone can account for the spectral structure and peak positions.

We have seen this broadening of the electron spectrum previously when including the inter-particle nuclear dynamics.
There, too, the electronic cross section is distributed over multiple vibronic transitions resulting in different outgoing electron energies.
Bound-bound transitions of the system produce discrete peaks in the spectrum, while a continuous line spectrum emerges from the dissociation of the system.\cite{jahr2025}

\subsubsection{Temperature dependence}

\begin{figure}[!ht]
    \centering
    \includegraphics[width=\linewidth]{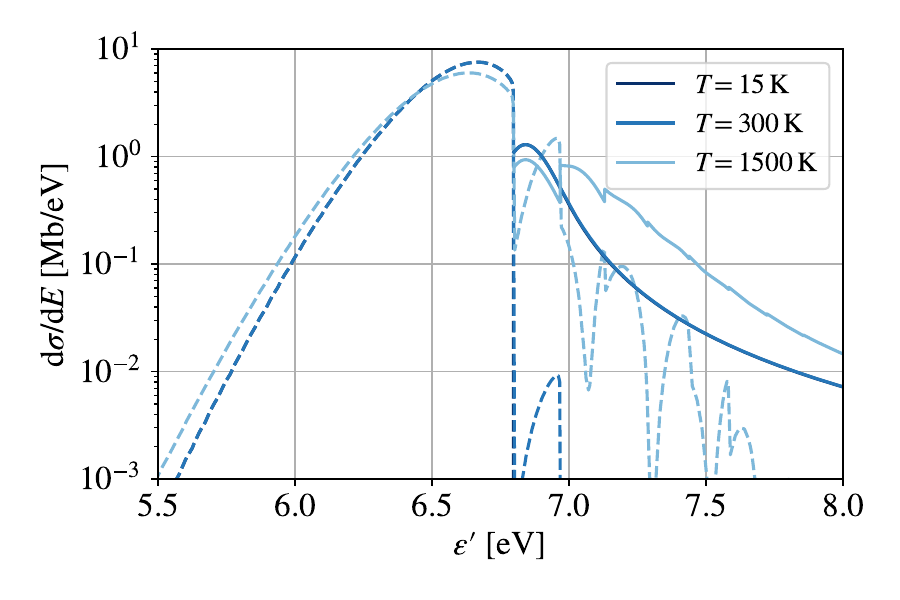}
    \caption{ICEC cross section at $\varepsilon=\SI{1}{eV}$ for \ce{H+} LiH vs. outgoing electron energy $\varepsilon'$ for different temperatures $T$ determined using the Franck-Condon model.
    Lighter shades indicate higher temperatures. 
    Solid: bound-bound transitions of LiH, folded with a Lorentz distribution; 
    dashed: bound-dissociative transitions.
    Additional features arise from increased thermal population of higher initial vibrational states.
    Figure taken from Ref.~\onlinecite{figshare_figures}.
    }
    \label{fig:boltzman}
\end{figure}

To model ICEC realistically, we incorporate vibrational averaging via the Boltzmann distribution in Eq.~\eqref{eq:boltzmann}, introducing temperature dependence into the electron spectrum shown in Fig.~\ref{fig:boltzman}.

We converted discrete peaks\cite{figshare_data} from bound-bound transitions of LiH to a continuous curve for easier comparison across different temperatures.
By folding the cross sections $\sigma(\varepsilon, \nu)$ in Eq.~\eqref{eq:boltzmann} with Lorentz (or Cauchy) distributions of width $2\gamma = \SI{0.16}{eV}$ at the peak positions $\varepsilon'_{\nu\to\nu_+}$
\begin{equation}
\begin{aligned}
    \sigma_L(\varepsilon, \nu) = \sum_{\nu_+} &\sigma(\varepsilon, \nu\to\nu_+) \\
    &\times \frac{\gamma/\pi}{(\varepsilon'-\varepsilon'_{\nu\to\nu_+})^2 + \gamma^2}
    \; H(\varepsilon'_{\nu \to E=0})
\end{aligned}
\end{equation}
we obtain an artificially continuous spectrum over all bound-bound transitions for each temperature.
A Heaviside step function $H(\varepsilon'_{\nu \to E=0})$ suppresses contributions in the outgoing electron energy range for respective bound-dissociative transitions of \ce{LiH+}. 

At low temperatures ($T=\SI{15}{K}$ and $\SI{300}{K}$) the electron spectra are nearly indistinguishable, except for a small bound-dissociative peak at $\SI{6.95}{eV}$, which only appears for a temperature of $\SI{300}{K}$.
At these temperatures the thermal population of the ground vibrational state $\nu=0$ of LiH exceeds $0.99$ and its cross section dominates in this temperature range.

The additional peak in the bound-dissociative curve originates from a peak of the bound-dissociative spectrum associated with $\nu=1$.
This peak is visible because the cross section from the ground vibrational state is zero above outgoing electron energies of $\varepsilon = \SI{6.8}{eV}$, as seen in Fig.~\ref{fig:spectrum-FC}.
Higher initial vibrational states decrease the ionization energy of LiH, allowing larger $\varepsilon'$.
This peak is one of the two peaks of the $\nu=1$ spectrum.
Due to the reflection principle, higher vibrational states produce an increasing amount of peaks in the bound-dissociative spectrum.
We have discussed this phenomena previously in the context of interparticle nuclear motion\cite{jahr2025}.

At $T = \SI{1500}{K}$, the maxima of the bound-dissociative and bound-bound lines are slightly reduced while the tails towards lower and higher electron energies $\varepsilon'$, respectively, are longer.
Both maxima are lower because the population of $\nu=0$ is reduced to 0.72, decreasing its contribution.
The population of excited initial vibrational states $\nu>0$ on the other hand increases, and their bound-bound transitions become relevant and raise the spectrum towards higher $\varepsilon'$.
Moreover, their longer bound-dissociative tails towards lower $\varepsilon'$ increase the spectrum at small $\varepsilon'$.
Additional peaks from $\nu>0$ emerge in the bound-dissociative spectrum towards higher $\varepsilon'$.

For systems with non-zero threshold for the kinetic energy of the incoming electron, an increase in temperature decreases the threshold energy.
Starting from higher initial vibrational states requires less electronic energy, and when temperature rises, their contributions to the cross section below the ground-vibrational threshold can become relevant.

Overall, increasing temperature further broadens the electron spectrum. 
Additional vibronic transitions from higher initial vibrational states of the molecules involved gain relevance both at the lower and upper energy end and introduce new features into the spectrum.

\section{Conclusion}
\label{sec:conclusion}

We extended the asymptotic model of ICEC to include the internal nuclear degrees of freedom of the electron acceptor and donor. 
Its power lies in the calculation of ICEC cross sections from properties of the individual species, such as their photoionization and photorecombination cross sections, thereby significantly reducing the computational cost and enabling the simulation of a greater variety of systems in the future.
In order to evaluate our expressions, we used and compared the impact of two different approaches:
(1) employing theoretical vibrationally resolved cross sections from the literature for the participating species, and
(2) using the Franck-Condon principle to model the vibronic transitions between Morse potentials fitted to potential energy curves of the species.

While theoretical vibrationally resolved cross sections can offer greater accuracy, the limited availability in the literature restricts their practical application. 
We have shown, that in such cases, the Franck-Condon principle remains a valuable alternative. 
Applying both approaches to proton ($\ce{H+}$) with lithium hydride (LiH), we observed qualitative agreement for bound-bound transitions in LiH. 
Moreover, the Franck-Condon principle allows to incorporate dissociative states in the description of the nuclear dynamics.
For the system investigated in this article, the Franck-Condon ICEC model therefore provided a more comprehensive description.

Our results on \ce{H+} LiH highlight that nuclear motion and temperature significantly alter the dynamics of the ICEC process. 
Dissociative pathways in LiH dominate over bound-bound transitions by an order of magnitude, underscoring the necessity of including dissociation of the molecules involved in ICEC.
Despite including nuclear motion, ICEC remains more probable than photorecombination of \ce{H+}.

A key finding is the strong dependence of the outgoing electron energy on internal nuclear dynamics.
At fixed incident electron energy, the electron spectrum features discrete peaks from bound–bound transitions and a continuous line from dissociation of LiH, contrasting with the single spectral line in the electronic case.

Increased temperatures further broadens the electron spectrum. 
Additional vibronic transitions from the molecules involved gain relevance both at the lower and upper electron energy end and introduce new features into the spectrum.

These findings emphasize the critical role of nuclear motion in ICEC, with implications for both theoretical modeling and experimental detection, particularly in weakly bound molecular systems.

\begin{acknowledgments}
We thank the DFG-ANR for financial support through the QD4ICEC project with grant number FA 1989/1-1.
E. F. furthermore acknowledges funding by LISA$^+$ at the University of Tübingen. 
\end{acknowledgments}

\appendix
\section{Appendixes}

\section*{Author statements}
\subsection*{Conflict of interest}
The authors have no conflicts to disclose.

\subsection*{Author Contributions}

\textbf{Conceptualization}: E.M.J., E.F.
\textbf{Formal Analysis}: E.M.J.
\textbf{Funding}: E.F.
\textbf{Investigation}: E.M.J.
\textbf{Methodology}: E.M.J., E.F.
\textbf{Software}: E.M.J.
\textbf{Supervision}: E.F.
\textbf{Visualization}: E.M.J.
\textbf{Writing -- original draft}: E.M.J.
\textbf{Writing -- review \& editing}: E.M.J., E.F.

\section*{Data availability}
The data that support the findings of this study are openly available and cited at the appropriate locations within this paper.

%\nocite{*}
\bibliography{refs}

\end{document}